\documentclass[preprint,12pt,longtitle,authoryear]{elsarticle}
\usepackage{subfigure}

\usepackage{amssymb,amsmath}

\journal{Computers \& Fluids}

\begin{document}

\begin{frontmatter}


%
%
%
%
%
%

\title{Radiative cooling in numerical astrophysics: the need for adaptive mesh refinement.}


\author[label1]{Allard Jan van Marle}
\author[label1,label2,label3]{Rony Keppens}
\address[label1]{Centre for Plasma Astrophysics, K.U. Leuven, Celestijnenlaan 200B, B-3001 Leuven, Belgium }
\address[label2]{FOM Institute for Plasma Physics Rijnhuizen, P.O. Box 1207
NL-3430 BE  Nieuwegein, the Netherlands}
\address[label3]{Astronomical Institute, Utrecht University, Budapestlaan 6
NL-3584 CD, Utrecht, the Netherlands}

\begin{abstract}
Energy loss through optically thin radiative cooling plays an important part in the evolution of astrophysical gas dynamics 
and should therefore be considered a necessary element in any numerical simulation.  
Although the addition of this physical process to the equations of hydrodynamics is straightforward, 
it does create numerical challenges that have to be overcome in order to ensure the physical correctness of the simulation. 
First, the cooling has to be treated (semi-)implicitly, owing to the discrepancies between the cooling timescale and the typical timesteps of the simulation. 
Secondly, because of its dependence on a tabulated cooling curve, the introduction of radiative cooling creates the necessity for an interpolation scheme.
In particular, we will argue that the addition of radiative cooling to a numerical simulation creates the need for extremely high resolution, 
which can only be fully met through the use of adaptive mesh refinement.
\end{abstract}

\begin{keyword}
Computational techniques: fluid dynamics \sep Radiative recombination 
 \sep Radiative transfer
in astrophysics \sep Stellar winds \sep Circumstellar envelopes
\PACS 47.11.-j \sep 78.60.-b \sep 95.30.Jx \sep 97.10.Me \sep 97.10.Fy
\MSC 68U20 \sep 76N15 \sep 76J20 \sep 85-08 \sep 85A25 \sep 85A30

\end{keyword}

\end{frontmatter}


\section{Introduction}
\label{sec-intro}
A gas at high temperature loses energy through radiation. 
In astrophysics the most common process to produce this radiation is through the recombination of ionized particles. 
This process can cause a significant decrease in local temperature over a short period of time, depending on the local conditions. 
In order to produce a correct numerical simulation of the evolution of a gas-structure on astrophysical scale, 
a method to quantify this form of radiative cooling has to be implemented.  
The most simple form is that of optically thin radiative cooling. 
Here it is assumed that the gas in which 
the photons are emitted is completely optically thin, so that any photon that is emitted 
will simply leave the physical domain of the simulation rather than being absorbed elsewhere. 
This is a valid approach for astrophysical phenomena, which tend to have low to extremely low densities. 
Indeed, the typical particle density of the interstellar medium is on the order of $10^{-24...-23}$~g/cm$^3$ (1...10 particles per cm$^3$), 
whereas the gas density of earth's atmosphere at sea level is approximately $1.2\times 10^{-3}$~g/cm$^3$.
The great advantage of the optically thin approach over a more complicated radiative transfer method 
is that it reduces the physics of radiative cooling to a purely local phenomenon. 
The energy of the gas in a given spot decreases, owing to its local density, temperature and degree of ionization. 
This maintains the localized nature of the Euler equations of gas dynamics (eq.~\ref{eq:euler}). 
Also, since the energy content of the escaping photons is lost to the gas anyway, there is no need to introduce an extra conserved variable for the radiation field internal energy.

The behaviour of an ideal gas can be described by solving the conservation equations for mass, momentum and energy here in Eulerian form \citep[e.g.][]{L98,C04,GKP10}
\begin{equation}
\begin{aligned}
\frac{\partial \rho}{\partial t}~+~\nabla\cdot(\rho\mathbf{v})~&=~0, \\
\frac{\partial (\rho\mathbf{v})}{\partial t}~+~\nabla\cdot(\rho\mathbf{vv})~+\nabla p~&=~\mathbf{F},\\
\frac{\partial e}{\partial t}~+~\nabla\cdot(e\mathbf{v}+ p\mathbf{v})~&=~G~+~\mathbf{v}\cdot\mathbf{F},
\label{eq:euler}
\end{aligned}
\end{equation}
with $\rho$ the mass density, $\mathbf{v}$ the velocity vector, $e$ the energy density, $p$ the thermal pressure and $\mathbf{F}$ and $G$ 
the momentum and energy source terms. 
These source terms can generally include such effects as radiation pressure, gravity etc. In our simulations, we will use optically thin radiative cooling as the only source term, making $\mathbf{F}=\mathbf{0}$ and $G$ as specified in eq.~\ref{eq:econs}. 
Since the pressure and density are related through 
\begin{equation}
e~=~\frac{1}{2}\rho v^2~+~\frac{p}{(\gamma-1)},
\end{equation}
with $\gamma$ the adiabatic constant, this set of equations forms a closed system that can be solved for a set of boundary values.

Optically thin radiative cooling \citep[e.g.][]{MB81,ML02,T09} is given by the equation
\begin{equation}
\frac{\partial p}{\partial t}~=~-(\gamma -1)n_i n_e \Lambda(T) \,\,\,\,\,\,\,[~\equiv~-(\gamma -1)\rho{\cal L}], 
\label{eq:decool}
\end{equation}
with $\partial p/\partial t$ the change in thermal pressure due to the cooling. 
$n_i$ and $n_e$ are the local ion particle density and electron density respectively, 
which relate to the total gas density through
$\rho=n_i m_i+n_e m_e$, with $m_e$ and $m_i$ the electron and ion masses.
$\Lambda(T)$ is a cooling function dependent on the local temperature $T$ and is usually taken from a table that is either constructed through observation, 
combining insights from astrophysics and laboratory experiments 
or through numerical simulations of the behavior of plasmas \citep[e.g.][]{DM72, MB81, ML02, S09}. 
In astrophysical literature one can find different tables, being constructed and gradually improved upon, that depend on the composition of the gas. The function ${\cal L}$ in the alternative way of writing the source term in eq.~\ref{eq:decool} specifies the energy losses per unit mass.

\subsection{Numerical challenges}
The cooling is implemented in a gas dynamics code by adding the effect of the right-hand term
in eq.\, \ref{eq:decool} to the Euler equation for conservation of total energy (eq.\ \ref{eq:euler}), 
which then includes an energy sink term $G$ and takes the form
\begin{equation}
\frac{\partial e}{\partial t}~+~\nabla\cdot (e{\mathbf
v})~+~\nabla\cdot (p{\mathbf v}) ~=~-n_i n_e \Lambda(T).
\label{eq:econs}
\end{equation} 
Though this is simple enough by itself, the unique nature of the radiative cooling function may lead to some difficulty with the implementation. 
Specifically, depending on the local characteristics of the gas, 
the cooling timescale
\begin{equation}
\tau_{\rm cool}~\sim~\frac{p}{n_i n_e \Lambda(T)},
\label{eq:tcool}
\end{equation}
can be much shorter than the dynamical timescale of the gas itself. 
Therefore, additional modifications to the original code are required in order
to either limit the timestep to the cooling timescale (Eq.\, \ref{eq:tcool}), 
or use an implicit calculation scheme for the cooling, even if the gasdynamics themselves are solved explicitly. 
The latter method, which has the advantage of stability inherent in implicit methods \citep{BVM01}, is usually preferred as the cooling timescales can become very short indeed,
which would slow down numerical simulations to an unacceptable level. Moreover, the typically tabulated nature of the cooling function $\Lambda(T)$ 
introduces the need for adequate interpolation formulae as well as potential difficulties across temperature ranges with strong variations (since the derivative ${\Lambda '(T)}$ is not known analytically).

\subsection{Astrophysical challenges}
Taking radiative cooling into account can change the result of a numerical simulation to a considerable extent. 
First of all, depending on circumstances, a large part of the available energy  can `leak' out of the gas through radiation. 
If this happens it can change the morphology of the gas, as areas of high thermal pressure disappear and the gas makes a transition from adiabatic to isothermal behavior.

A second effect is the sudden formation of radiative cooling instabilities. 
These can result from either the density or the temperature dependence of the cooling. 
In the case of density dependent cooling instabilities, high density regions lose more energy than their surroundings, and therefore have a lower temperature, which leads to a lower local pressure. 
Hence, these regions will be compressed, causing an increase in local density, which in turn means a further increase in the cooling rate. 
This process can repeat itself until the temperature of the gas drops to zero, or whatever floor temperature has been set for the simulation, depending on the physical conditions one tries to simulate. 
As a result, simulations that include radiative cooling tend to show high density regions of much smaller physical size and higher local density than if the same model had been run without radiative cooling. 

Thermal instabilities can also result under isobaric conditions, in those temperature ranges where $\Lambda(T)$ decreases for increasing $T$. 
A runaway condensation results when 
\begin{equation}
\biggl[\frac{\partial {\cal L}}{\partial T}\biggr]_p~=~\biggl[\frac{\partial {\cal L}}{\partial T}\biggr]_\rho~-~ \frac{\rho}{T}\biggl[\frac{\partial {\cal L}}{\partial \rho}\biggr]_T~<~0,
\end{equation}
with ${\cal L}$ the energy gain/loss per unit mass due to non-adiabatic processes. 
Similarly, sound waves can turn into overstable modes under isentropic conditions $[\partial {\cal L}/\partial T]_S <0$ 
\citep{F51,P53}, where $S=p\rho^{-\gamma}$ denotes the entropy.

Moreover, the high density structures formed by the cooling can also serve as seed for other instabilities, such as Kelvin-Helmholtz (when two parallel flows have a strong shear \citep{C81}), Rayleigh-Taylor (when gravitational, centrifugal or thermal pressure driven acceleration occurs \citep{C81, FGR02}) and thin shell instabilities (which are the result of a thin shell being compressed between two areas with asymmetric pressure gradients \citep{V83}).
As we will demonstrate in this paper, resolving such high density structures presents a challenge to the numerical code and necessitates an increase in resolution, 
which can best be achieved through the use of adaptive mesh refinement. 

Finally, radiative cooling allows us to compare our simulations directly with observations, as
the radiative flux produced in this way is an observable quantity.

\begin{table}
\label{tab:input}
\caption{Physical input parameters}
\begin{tabular}{p{0.45\linewidth}ll}
\hline
$\frac{dM}{dt}$             &  $10^{-6}\, \mathrm{M}_\odot$/yr & $(=6.3\times 10^{19}\, \mathrm{g/s})$ \\
$V_\infty$                        &  $1500\, \mathrm{km/s}     $            &     \\
$\rho_{\mathrm{ISM}}$ & $10^{-22.5}\, \mathrm{g/cm^3}$ & \\
$T_{\mathrm{ISM}}$ & 100\, K & \\
\hline
\end{tabular}
\end{table}

\section{Astrophysical application: stellar wind expansion}
\label{sec-astroprob}
As our test case, we use the expansion of a supersonic stellar wind into a surrounding constant density interstellar medium (ISM). 
This problem was approximated analytically by \citet{CMW75}, \citet{Wetal77}, \citet{OM88} and others
and has since then been tested numerically by e.g. \citet{GML96, GLM96}, 
with the numerical results showing close qualitative and quantitative agreement with the original analytical model.

Schematically, the wind blown bubble shows the following structure. 
The stellar wind collides with the ISM. 
As the wind slows down because of the collision, its kinetic energy is converted to thermal energy, creating a `hot bubble' of shocked wind material, 
which is contained by the reverse shock (R1) on one end and a contact discontinuity (R2) on the other. At the contact discontinuity (R2), 
the high thermal pressure of the `hot bubble' allows it to expand outward into the ISM, sweeping up a shell of shocked ISM (see fig.~\ref{fig:nocool_rho_temp} and further).
According to the analytical model the outer radius of the shocked ISM shell should be at time $t$ at position:
\begin{equation}
R_3~=~\biggl(\frac{250}{308\pi}\biggr)^{1/5} L_w^{1/5} \rho_{\mathrm{ISM}}^{-1/5} t^{3/5},
\label{eq:weaver}
\end{equation}
with $L_w=\frac{1}{2}\frac{dM}{dt}V_\infty^2$ the mechanical luminosity of the wind with $dM/dt$ the mass loss rate, $\rho_{\mathrm{ISM}}$ the density of the ambient interstellar medium, $V_\infty$ the terminal velocity of the wind and $t$ the time \citep{Wetal77}. The numerical factor is a result of the choice of units, in this case cgs (centimeter, gram, second).

It should be noted here that this model is somewhat simplified from an astrophysical point of view, 
as it assumes that the stellar wind expands into a cold medium. 
More recent models \citep{FHY03,FHY06, VLG05,VLG07}, 
which include the effect of photo-ionization by radiation from the progenitor star show a rather more complicated result. 
In fact, even these models neglect external factors such as stellar winds and ionizing radiation from neighboring star systems. 
Such factors cause inhomogeneities in the ISM, which complicate the matter further
\citep{Metal06}. 
Nevertheless, this is a good test case, 
since it is extremely well documented and the result can change considerably due to the influence of radiative cooling. 
For our input parameters we choose the parameters shown in Table 1. 
The mechanical luminosity ($L_w$) which follows from these parameters is high, 
with the mass loss rate $dM/dt$ and wind velocity $V_\infty$ reflecting values typically found only in extremely massive stars ($\geq\, 60\, M_\odot$) of O and hydrogen rich Wolf-Rayet type \citep{Letal94, GML96, VLG07}. 
These values were chosen deliberately to create a powerful, high density shock which causes a high cooling rate.

\begin{figure*}
 \centering
\resizebox{\hsize}{!}{\includegraphics[width=0.95\textwidth]{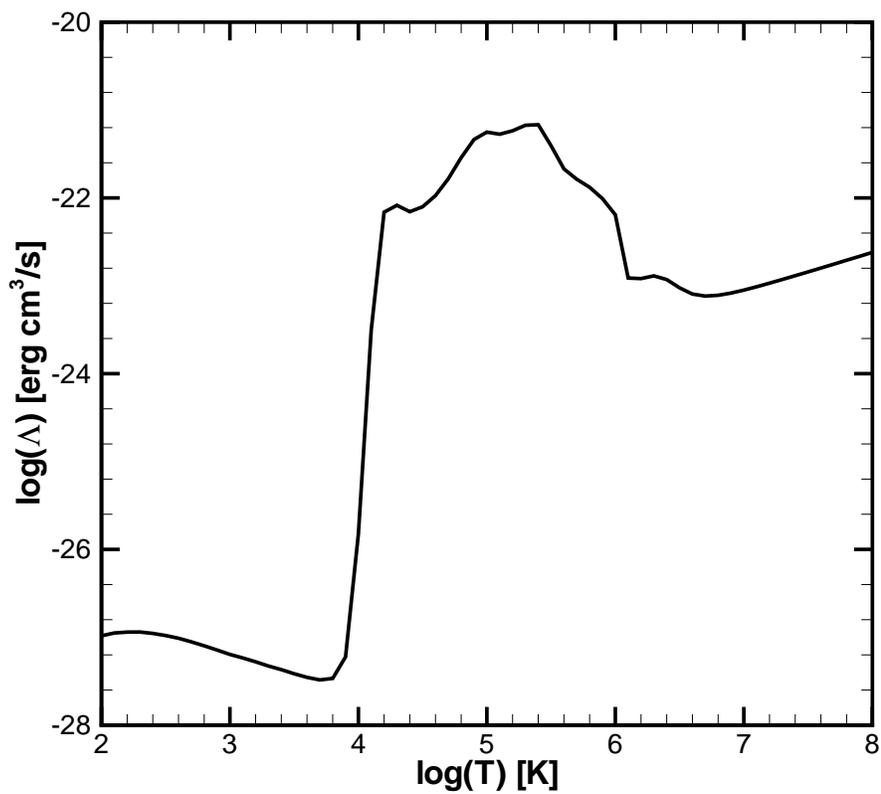}}      
\caption{Cooling table for gas at solar metallicity according to \citet{ML02}. Note the steep jump in cooling rate around $10^4~K$, which corresponds to the (de-)ionization of hydrogen.}
 \label{fig:cooltable}
\end{figure*}

\section{Numerical approach}
\label{sec-numerical}
For our simulations we use the AMRVAC code \citep{Ketal03, amrvac07}. 
This is a fully conservative code, which can solve e.g. the Euler equations of hydrodynamics on a grid
that can be either Cartesian, cylindrical or spherical. 
The code includes an adaptive mesh refinement (AMR) scheme to adjust the resolution depending on preset criteria. 
In addition, AMRVAC allows the use of a wide variety of shock capturing solvers. 

\subsection{Discretization of the Euler equations}
\label{sec-num_eul} 
The numerical form of the Euler equations, as used in the AMRVAC code has been described before \citep[e.g.][]{Ketal03}, but we briefly recall the main solver strategy as employed for the simulations reported here. 
The integration scheme advances conserved variables (mass, momentum and total energy), which represent grid cell values in the standard finite volume sense. 
The numerical update takes the general form (specified to 1D Cartesian for simplicity)
\begin{eqnarray}
\mathrm{predictor} & \,\,\,\,\,\, & U_i^{n+\frac{1}{2}}~=~U_i^{n} ~-~\frac{\Delta t}{2\Delta x} \biggl( F(U^{L,n}_{i+1/2})- F(U^{R,n}_{i-1/2}) \biggr), \nonumber \\
\mathrm{corrector} &  \,\,\,\,\,\, & U_i^{n+1}~=~U_i^{n} ~-~\frac{\Delta t}{\Delta x} \biggl( F_{i+1/2}^{LF,n+\frac{1}{2}} - F_{i-1/2}^{LF,n+\frac{1}{2}} \biggr),
\end{eqnarray}
with $U_i$ the set of conserved variables $(\rho, \rho v, \rho v^2/2 +p/(\gamma-1))$ at the cell center of grid $i$, and $F$ the physical fluxes deduced from eq.~\ref{eq:euler} to be $(\rho v, \rho v^2 + p, ev+pv)$, which are in the (non-conservative) Hancock predictor step evaluated at cell interface states $U^{L,n}_{i+1/2}, U^{R,n}_{i-1/2}$. In the corrector step, we use the conservative
Total Variation Diminishing Lax-Friedrichs
(TVDLF) \citep{TO96,Y89} spatial discretization. 
This method specifies the fluxes  $F_{i+1/2}^{LF}$ according to
\begin{equation}
\begin{aligned}
F_{i+1/2}^{LF}~&=~\frac{1}{2}\biggl[F(U^L_{i+1/2}) + F(U^R_{i+1/2}) \\
                   &-~|c^{\mathrm{max}}\biggl(\frac{U^L_{i+1/2}+U^R_{i+1/2}}{2}\biggr)| \biggl(U^R_{i+1/2} - U^L_{i+1/2}\biggr)\biggr].
\end{aligned}
\end{equation}
The locally computed $c^\mathrm{max}$ denotes the maximum physical propagation speed at the (averaged) cell interface state, and is for hydrodynamics given by $|v|+\sqrt{\gamma p/\rho}$.
To obtain the cell interface states from the cell center values, a limited linear reconstruction determines the
$U^L_{i+1/2}$ and $U^R_{i+1/2}$ states as
\begin{equation}
\begin{aligned}
U^L_{i+1/2}~&=~ U_i +\bar{\Delta U_i}/2,  \\
U^R_{i+1/2}~&=~ U_{i+1} - \bar{\Delta U_{i+1}}/2 ,
\end{aligned}
\end{equation}
which involves a limited slope $\bar{\Delta U_i}=\Delta U_i \phi(r_i)\equiv \Delta U_i \phi(\Delta U_{i-1}/\Delta U_i)$.
The limiter is here written to act on the cell differences $\Delta U_i=U_{i+1}-U_i$, but in AMRVAC these can also be employed on the corresponding primitive variables $(\rho, v, p)$ or other user selected (e.g. logarithmically stretched) combinations. Many flavors are implemented, but for most of our calculations we use the `minmod' flux limiter, which gives for $\phi(r)$ the behavior
\begin{equation}
\phi(r)=
 \begin{cases}
 1& \text   {if $r\geq 1$},\\
r& \text  {if $0\geq r \geq 1$},\\
0& \text  {if $r<0$}.
 \end{cases}
\end{equation}
To compare the effect of different flux limiters on the result, we also run simulations with the `van Leer'~ \citep{L74} limiter
\begin{equation}
\phi(r)=
 \begin{cases}
\frac{2r}{1+r} & \text   {if $r\geq 0$},\\
0& \text  {if $r<0$}.
 \end{cases}
\end{equation}
Since our astrophysical problem involves density and pressure jumps of several orders of magnitude, 
we use the logarithm of these positive primitive variables in the limiter.

In order to avoid numerical instability in this explicit two-step integration procedure, the size of each timestep has to be limited by the Courant-Friedrichs-Levy (CFL)
condition, which stipulates that the time step must obey 
\begin{equation}
\Delta t^n ~\leq~\biggl[\frac{\Delta x}{c^{\mathrm{max}}}\biggr]_i^n.
\label{eq:CFL}
\end{equation}
Here $\Delta x$ is the (local when AMR) size of a gridcell. The current AMRVAC code uses a single timestep for the entire grid hierarchy in the AMR, so the righthand value of Eq.~\ref{eq:CFL} is calculated for all gridcells to find the global minimal value. 
The new timestep becomes a fixed fraction of this value, in our case 0.25. In the simulations discussed below, this scheme is only slightly modified to handle the actual spherical geometry, and multi-dimensional simulations handle fluxes from multiple directions in unsplit fashion. 

\subsection{Radiative cooling methods}
In order to simulate the effect of radiative cooling, an extra module has been added to the code, 
which updates the local energy of the gas according to Eq.~\ref{eq:decool}. 
This new module allows for various approaches to handle the sink term.
The simplest method is a fully explicit cooling routine:
\begin{equation}
e^{n+1}~=~e^{n} - n_i^{n} n_e^{n} \Lambda(T^{n}) \Delta t^n,
\label{eq:cexplicit}
\end{equation}
where ${n}$ and ${n+1}$ denote discrete time levels before and after the timestep $\Delta t^n$ respectively. 
For this approach, numerical stability is maintained by forcing the timestep $\Delta t^n$ to remain
below a pre-set fraction of the cooling timescale (Eq.\, \ref{eq:tcool}) everywhere in the
grid, this upper limit being enforced on top of the standard CFL condition.

Alternatively, the cooling can be calculated using a semi-implicit scheme so that: 
\begin{equation}
e^{n+1}~=~e^{n} - n^n_i n^n_e \Lambda(T^{n+1}) \Delta t^n.
\label{eq:cimplicit}
\end{equation}
This way, no extra limit has to be placed on the size of the timestep, but computationally it is 
expensive, since the numerical scheme must estimate the cooling rate at
$T^{n+1}$. 
In our implementation the cooling value $\Lambda(T^{n+1})$ is estimated through a half-step refinement routine 
that iterates until a preset precision is achieved. 
This is a reliable method but the number of iterations can become large (typically about 10...20 iterations to reach a relative precision of $10^{-5}$).

Several hybrid solutions have also been implemented. 
One uses a cooling value found by taking the average 
between $\Lambda(T^n)$ and $\Lambda(T^{n+1})$, without adjusting the size of the timestep, 
while the second allows the timestep to be split up into smaller, explicitly treated `cooling
steps'. 
The first has the advantage of providing a very fast solution at the expense of physical 
accuracy, whereas the latter is more precise, but can be nearly as slow as a fully explicit
method, depending on the nature of the simulation.

Finally, we have implemented the new exact-integration method proposed by \citet{T09}. 
This method uses an exact integration of the cooling equation \ref{eq:decool}. 
The temperature evolution due to radiative cooling of a gas parcel at constant density 
is calculated in advance, starting at extremely high temperature (usually $10^8$K) and followed all the way down to the lowest
temperature (usually $10^2$K) for which a cooling value is known. 
Since this is only done once and does not have to be repeated during the actual simulation, it
can be done in very small steps to increase accuracy. 
Once the actual simulation starts, the code evaluates for each grid point where the local gas
is on this temperature evolution and where it should end up for a given timestep. 
It then updates the temperature accordingly so that
\begin{equation}
T^{n+1}~=~ Y^{-1}\biggl[Y(T^n) + \frac{T^n}{T_{ref}} \frac{\Lambda(T_{ref})}{\Lambda(T^n)}
\frac{\Delta t^n}{\tau_{cool}}\biggr],
\label{eq:cexact}
\end{equation}
where $Y$ is the dimensionless temperature evolution function of a single gas element starting at the maximum temperature of the cooling table as obtained a priori. 
In this formula, $T_{ref}$ is an arbitrarily
chosen reference temperature (usually the highest temperature in the cooling table, though this isn't requisite). 
This method has proved to be both faster and more reliable than either explicit or implicit
methods and is the one that we selected for our simulations.
For the cooling curve we use the cooling table calculated  by \citet{ML02} for gas of solar metallicity. 
The actual (tabulated) relation between $\Lambda$ and $T$ is shown in  a log-log plot (fig.~\ref{fig:cooltable}), demonstrating 
the extreme challenges due to the large (several orders of magnitude) differences.

The ion density $n_i$ and electron density $n_e$ follow from the mass density and the composition of the gas. 
For simplicity we assume that hydrogen completely dominates the gas (a reasonable assumption for most astrophysical situations) and that the gas that actually cools is fully ionized. Therefore, $n_i=n_e=\rho/m_h$,
with $m_h=1.67\times10^{-24}$g the hydrogen mass.

\subsection{Grid setup and AMR strategy}
In order to achieve a high resolution in those areas where it is necessary, we use adaptive mesh refinement (AMR). 
The AMR scheme used in the AMRVAC code is a parallellized, modern block-tree variant of the original scheme described by \cite{BC89}. 
This involves the dynamical generation (and destruction) of hierarchically nested grids of a fixed subgrid (block) size, up to a preset maximum level. This maximum level, along with the corresponding effective resolution obtained, is reported for all cases studied below.
Whether to create or destroy a grid at any given level is decided automatically at each timestep based on the \citet{L87} prescription. 
This estimates a weighted 2$^\mathrm{nd}$ derivative of a variable $w$ in grid point $i$ from
\begin{equation}
\sqrt{\frac{\Sigma_{i_1}\Sigma_{i_2} \biggl(  \Delta x_{i_1}\Delta x_{i_2}\biggl[\frac{\partial^2 w}{\partial x_{i_1}\partial x_{i_2} }\biggr]_i\biggr)^2}
{\Sigma_{i_1}\Sigma_{i_2}\biggl(|\Delta x_{i_1} \biggl[\frac{\partial w}{\partial x_{i_1}}\biggr]_{i-1}|+|\Delta x_{i_1} \biggl[\frac{\partial w}{\partial x_{i_1}}\biggr]_{i+1}| + f|\tilde{w}|\biggr)^2}}.
\label{eq:lohner}
\end{equation}
This formula already applies to multi-dimensional simulations, and has a wavefilter parameter $f$ which we fix to $10^{-2}$. The $\tilde{w}$ indicates an average over all neighboring grid cells in directions $i_1, i_2$. This formula is actually used on a user-selected set of variables $w$ from the set of conserved variables $U$.
In our case, we have chosen the density as the variable to initiate refinement. 
Since our simulations are primarily concerned with resolving a high density feature (the circumstellar shell), 
this is the most logical choice.
While the formula~\ref{eq:lohner} quantifies the local error for variable $w$ in grid point $i$, the automated block-based regridding works as follows. A fixed gridblock between levels 1 and the maximal level allowed is refined if any point on it has this local error exceed a pre-set maximum tolerance value set by the user (0.1 in all our simulations). In contrast, if all points on the fixed grid-block have their error below a fraction (1/8 for our runs) of this maximum tolerance, the grid is coarsened. This procedure is then complicated by the proper nesting criterion, ensuring no grid changes by more than 1 level across bounding grid blocks. 

\subsection{Initial and boundary conditions}
 The simulation is set up by using a 1-D spherically symmetric grid with 400 gridpoints to cover a physical distance starting at 
$10^{17}$~cm and ending at $10^{19}$~cm. 
Density and pressure in the grid are taken to be constant, with the gas velocity set to zero.  
Boundary conditions at the outer radial boundary are set to continuous to allow a free outflow. 
At the inner boundary, conditions are set to simulate the inflow of a steady stream of matter according to the values in Table~1. This corresponds to employing standard Dirichlet boundary conditions at the inlet. As usual in any finite volume code, the boundary conditions include filling of ghost cells, and the stencil of our second order shock-capturing scheme used demands the use of two ghost cell layers.
Unlike e.g. \citet{SRT09} we do not attempt to simulate the acceleration zone of the wind. 
Instead, we assume that the wind has already reached its terminal velocity ($V_\infty$) by the time it crosses the inner boundary of our grid.
The outer boundary is set to continuous (i.e. zero gradient extrapolation), 
though this is by and large irrelevant as the simulation is stopped before the expanding wind-bubble reaches the outer boundary. 

For our first simulation we use a fixed grid of 400 cells without refinement (AMR~1). 
Subsequently, the simulation is repeated with increasing levels of refinement (AMR~2 through AMR~11). 
Each level of refinement corresponds to a doubling of the effective resolution so that AMR~11 corresponds to a static grid of
409\,600 grid points.
To show the effect of radiative cooling on both the physical result as well as the numerical simulation we do this 
for the same simulation both with and without radiative cooling.
We also include runs with different numerical solvers (van Leer flux-limiter and explicit and semi-implicit radiative cooling) to show how they influence the result. 

Finally, we show the results of a single 2-dimensional run to demonstrate how shifting to higher dimensions changes the result due to local instabilities and to give the reader an idea of  what an actual circumstellar shell would look like. For this run we use a van Leer flux-limiter, which gives better results at higher resolution, with 4 levels of adaptive mesh (as in the 1-D AMR~4 simulation). The base grid has 400 radial gridpoints and 200 azimuthal gridpoints covering the same radius as the 1-D simulations and an azimuthal opening angle of 90$^{\mathrm{o}}$.

\subsection{Computational resources}
All calculations were done on the Vic3 supercomputer at K.U.Leuven, Belgium, which originally consisted of a 928 core cluster (L5420 CPUs, 1GB RAM/core) from SGI, recently updated with 640 cores (Xeon 5560 CPUs, 3GB RAM/core). All computational nodes are connected through a double DDR infiniband network. 
We typically use a single node of 8 CPUs for the 1-D simulations with execution times varying from about 2 minutes for the low levels of resolution to a maximum of about 24 hours for the largest simulations.

\begin{figure*}
 \centering
\resizebox{\hsize}{!}{\includegraphics[width=0.95\textwidth]{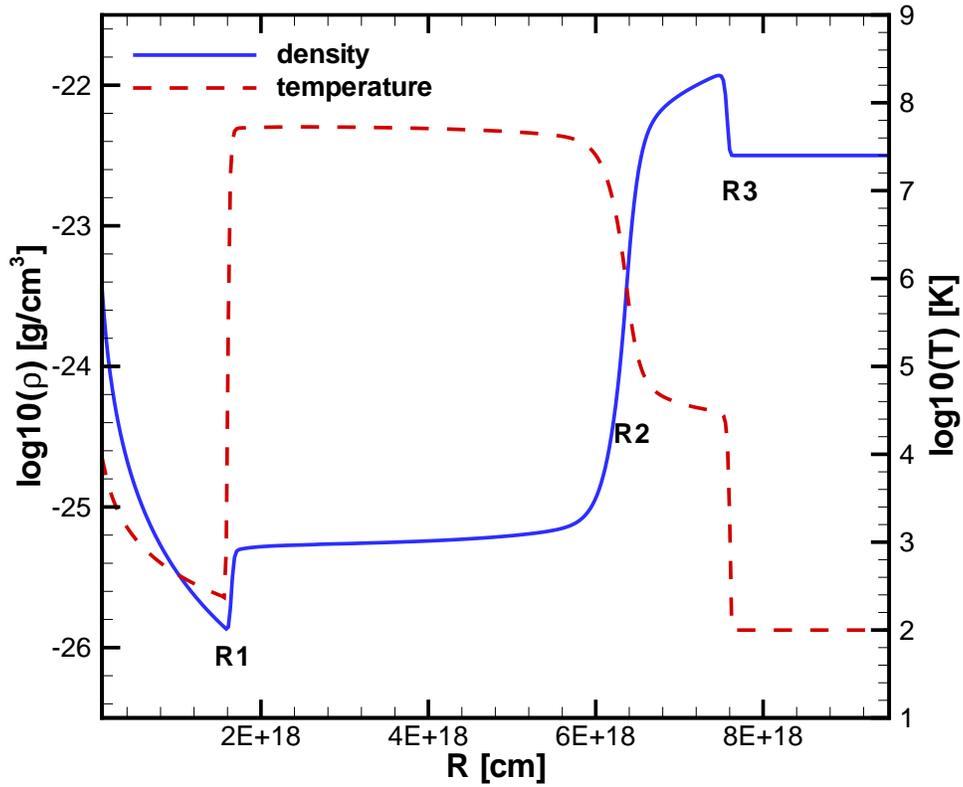}}      
\caption{This figure shows density and temperature after $1.25\times10^{12}$~seconds for a fixed grid simulation with 400 grid points. The wind termination shock (R1) is well resolved. The forward shock (R3) somewhat less so and the contact discontinuity (R2) clearly needs a higher resolution.}
 \label{fig:nocool_rho_temp}
\end{figure*}

\begin{figure*}
 \centering
\mbox{
\subfigure
{\includegraphics[width=0.5\textwidth]{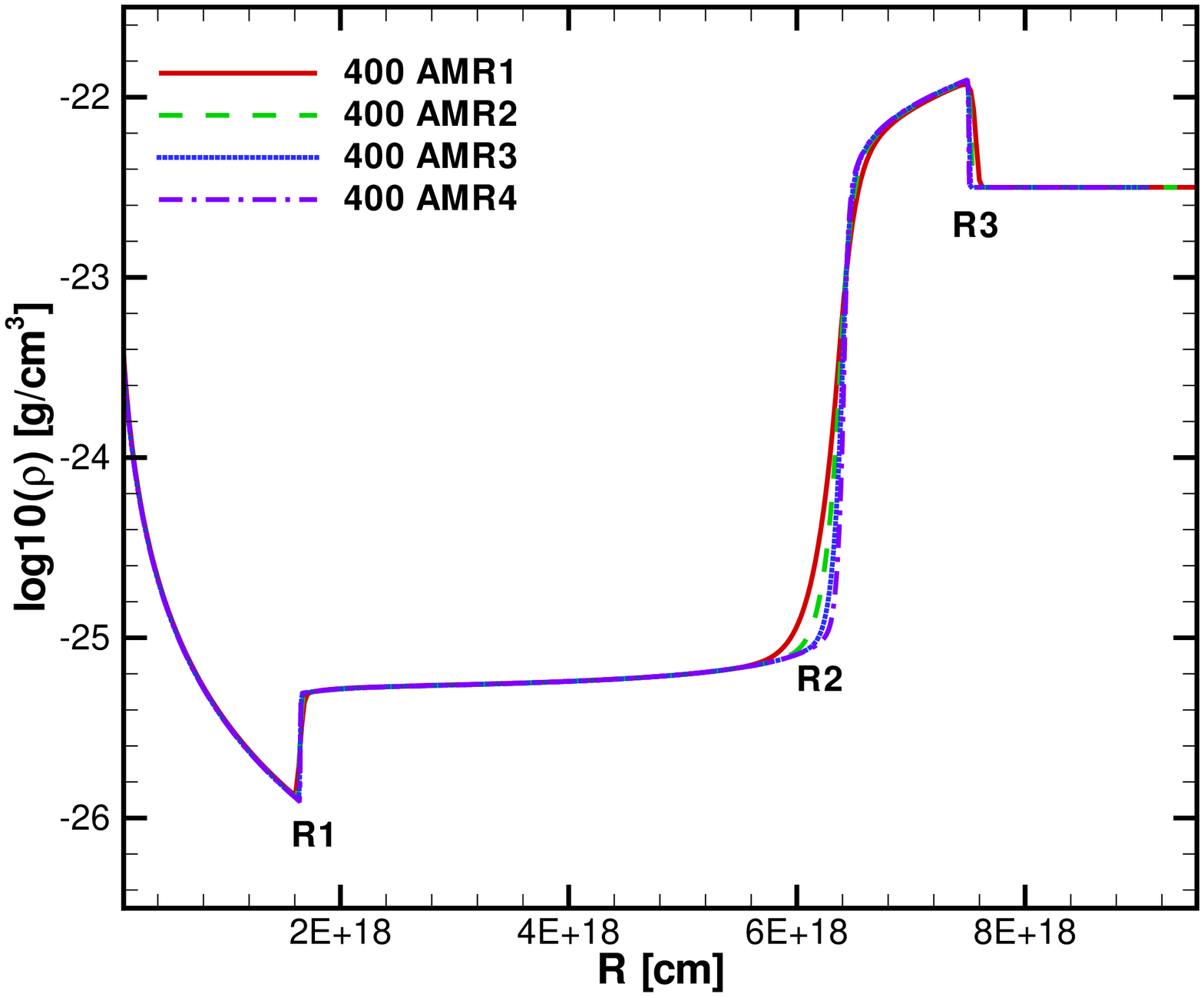}}
\subfigure
{\includegraphics[width=0.5\textwidth]{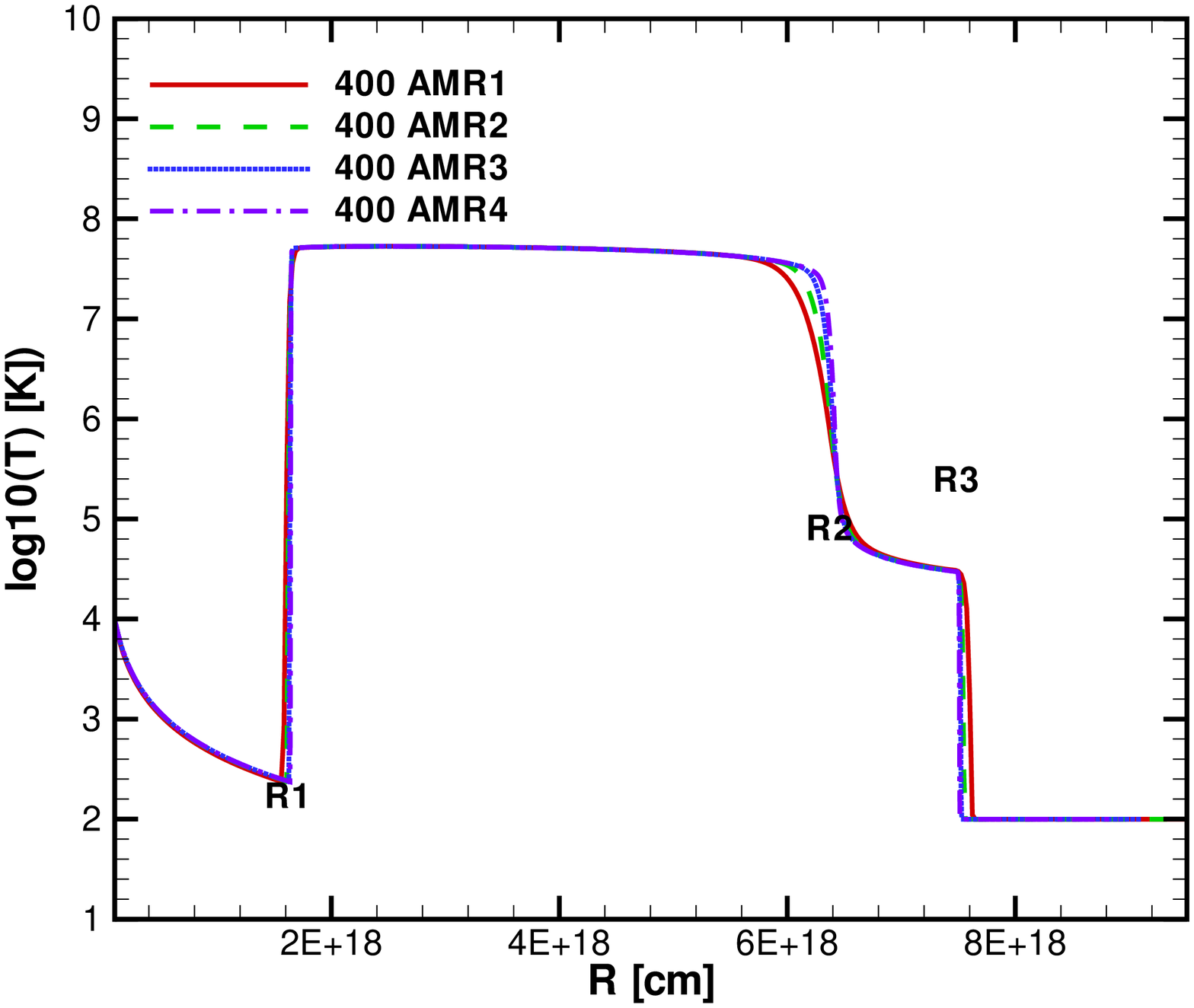}}}
\caption{Results for simulations without radiative cooling, showing density (left) and temperature (right) after $1.25\times10^{12}$~seconds (as in fig.~\ref{fig:nocool_rho_temp}). Each simulation has a different number of refinement levels, starting from 400 gridpoints with no refinement (AMR1) and ending with 4 levels (3 levels on top of the original grid) (AMR4). Note that the position of the shell does not change. Only the shape gradually converges, with the contact discontinuity R2 most affected by the increased resolution.}
 \label{fig:nocool}
\end{figure*}

\section{Results}
\label{sec-results}
All simulations described use the setup mentioned earlier.
We run simulations both with and without radiative cooling and 
compare the results of a number of different cooling methods and numerical solvers. We do simulations with different maximal grid levels (with a factor 2 refinement between consecutive levels), compare to uniform grid runs, and vary the physical parameters as well.

\subsection{No cooling}
We start with a series of simulations which were run without radiative cooling. 
The results are presented in figs.~\ref{fig:nocool_rho_temp} and \ref{fig:nocool}.
Figure~\ref{fig:nocool_rho_temp} shows the mass density and temperature of the gas at time~$1.25\times10^{12}$~seconds for a simulation with a fixed grid of 400 points. 
The wind termination shock (R1) and forward shock (R3) are already quite well resolved, 
although the resolution of the forward shock in particular leaves room for improvement. 
The contact discontinuity (R2) is less well resolved and shows the need for a higher effective resolution.
This plot clearly follows the shape predicted by \citet{Wetal77}, with a free-streaming wind coming from the star, then a layer of (near-)constant density shocked wind material (between R1 and R2) and finally a swept up shell expanding into the surrounding interstellar medium.  
According to the analytical solution by \citet{Wetal77}, the outer edge of the shell should be at $6.47\times10^{18}$~cm for these input parameters. 
In the simulation the outer edge of the shell clearly lies at a larger distance.
However, the analytical solution depends on the shell being thin compared to the shocked wind layer that pushes it outward. 
In the numerical solution this is obviously not the case due to the internal gas-pressure of the shell, which \citet{Wetal77} neglect.

Figure~\ref{fig:nocool} shows the same solution as fig.~\ref{fig:nocool_rho_temp} for four simulations with increasing number of levels of adaptive mesh. 
The effect of increased resolution is small. 
Both shocks and contact discontinuities become sharper, but the shape of the bubble remains the same and the values of density and temperature are generally the same, except in the transitional regions. 
The simulations with 3 and 4 levels of refinement  (AMR~3 and AMR~4 respectively) overlap each other almost completely, such that a further increase in resolution is unnecessary when pure Eulerian dynamics is studied.  

\begin{figure*}
 \centering
\resizebox{\hsize}{!}{\includegraphics[width=0.95\textwidth]{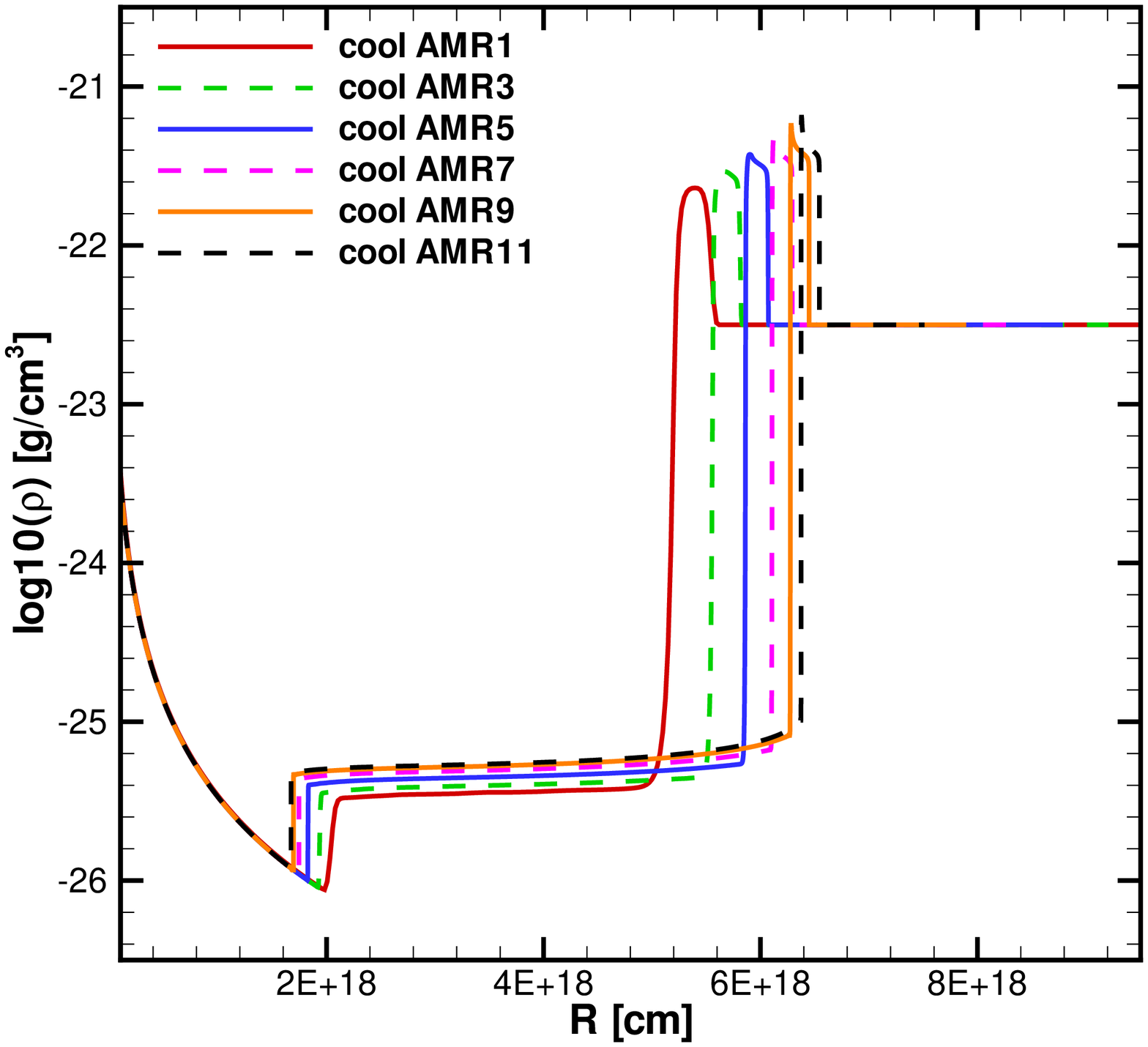}}      
\caption{Similar to fig.~\ref{fig:nocool}, but with radiative cooling. The shell is much thinner due to compression. For low resolution the shell is poorly resolved, leading to considerable numerical errors. Only for high resolutions, from about AMR~9 onwards, do the results converge. We only show the simulations with odd numbered levels of refinement to make the figure clearer.}
 \label{fig:cool1}
\end{figure*}

\begin{figure*}
 \centering
\resizebox{\hsize}{!}{\includegraphics[width=0.95\textwidth]{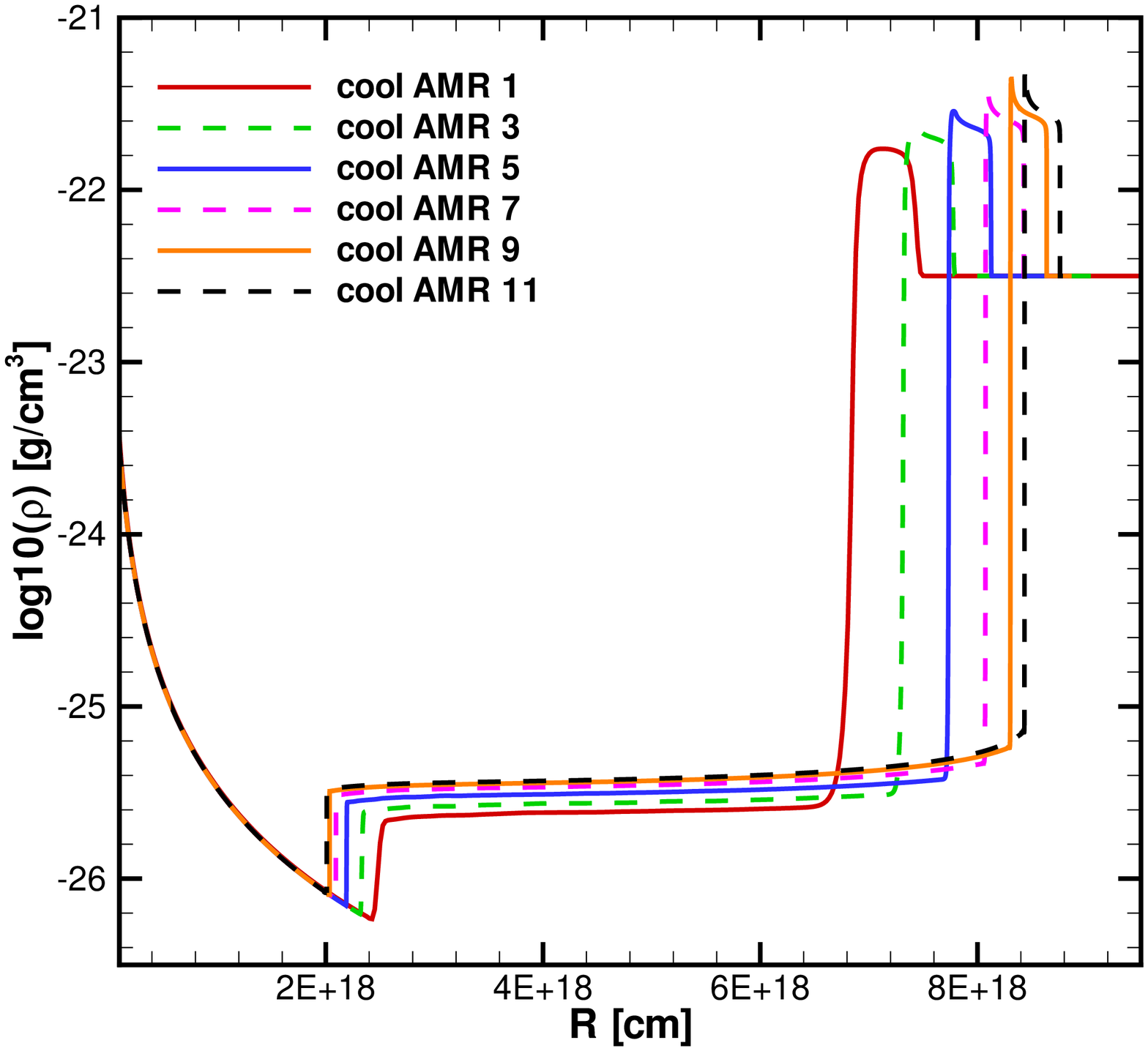}}      
\caption{Similar to fig.~\ref{fig:cool1}, but after $2.0\times10^{12}$~seconds. The pattern is the same. At high resolution the difference between simulations 
is largely irrelevant, since the absolute difference between shell positions stays approximately the same, while the total size of the circumstellar bubble increases.}
 \label{fig:cool2}
\end{figure*}

\subsection{With cooling}
The results of our simulations with cooling are given in figs.~\ref{fig:cool1} and \ref{fig:cool2}, which show the density of the gas at time $1.25\times10^{12}$  (as in fig.~\ref{fig:nocool}) and time $2.0\times10^{12}$~seconds.
In fig.~\ref{fig:cool1} the simulation without refinement shows clearly that the resolution of the grid is too low for an accurate result, 
as the swept-up shell is not properly resolved, looking more like a Gaussian curve than the straight angles that should occur at both shocks and discontinuities. 
Only at four levels of refinement and higher the simulations start to give a much better result, 
showing the top of the shell as a flat plateau, though the edges are still too rounded and no internal structure in the shell is visible. 
This is solved by adding more resolution. 
Internal structure of the shell only appears at very high levels of resolution ($>6$). 
This structure is a direct result of the radiative cooling instability: high density regions cool more rapidly and therefore have a lower thermal pressure. As a result they get compressed by the surrounding gas, increasing their density even further, which in turn will cause them to cool more rapidly. 

Figure \ref{fig:cool2} shows the same simulations, but at a later time ($2.0\times10^{12}$~seconds).  
As time progresses, the amount of mass swept-up in the shell increases, which in turn causes the shell to become thicker. 
As a result it becomes easier to resolve. 
By now even the first simulation is beginning to resolve the shell and the 4 level simulation (AMR~4) is already showing internal structure. 
The three highest resolution shells are very similar and overlap each other almost completely. 
Still, the difference in location between the shells at low and high refinement shows the need for high effective resolution. 
When comparing figs.~\ref{fig:cool1} and \ref{fig:cool2} for the models with the highest levels of refinement we note 
that the absolute difference does not change over time. The slight difference actually originates during the early phases of the simulation. 
By starting the simulation with a direct interaction between wind and stationary medium, we created a situation where the initial shocked gas layer will be very thin and even a very fine mesh will only be an approximation. Only once the 
shocked gas layer is properly resolved, does the difference in resolution no longer matter. 

\begin{figure*}
 \centering
\resizebox{\hsize}{!}{\includegraphics[width=0.95\textwidth]{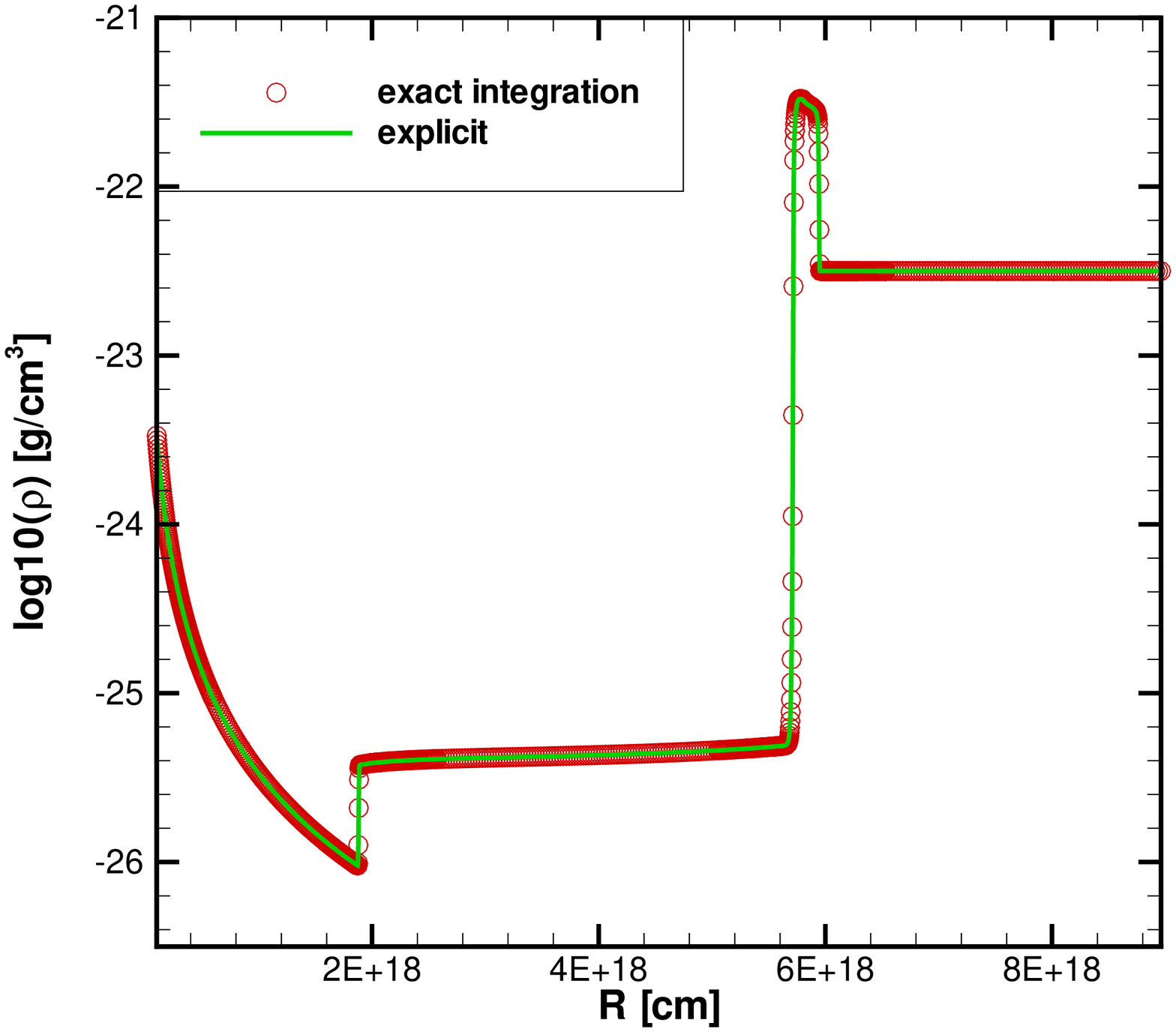}}      
\caption{The same point in time as in figs.~\ref{fig:nocool} and \ref{fig:cool1} for two simulations both with 4 levels of refinement. One using the exact integration method for the cooling and the other using a fully explicit scheme with limited time steps. The results are indistinguishable.}
 \label{fig:2coolmethods}
\end{figure*}

\begin{figure*}
 \centering
\resizebox{\hsize}{!}{\includegraphics[width=0.95\textwidth]{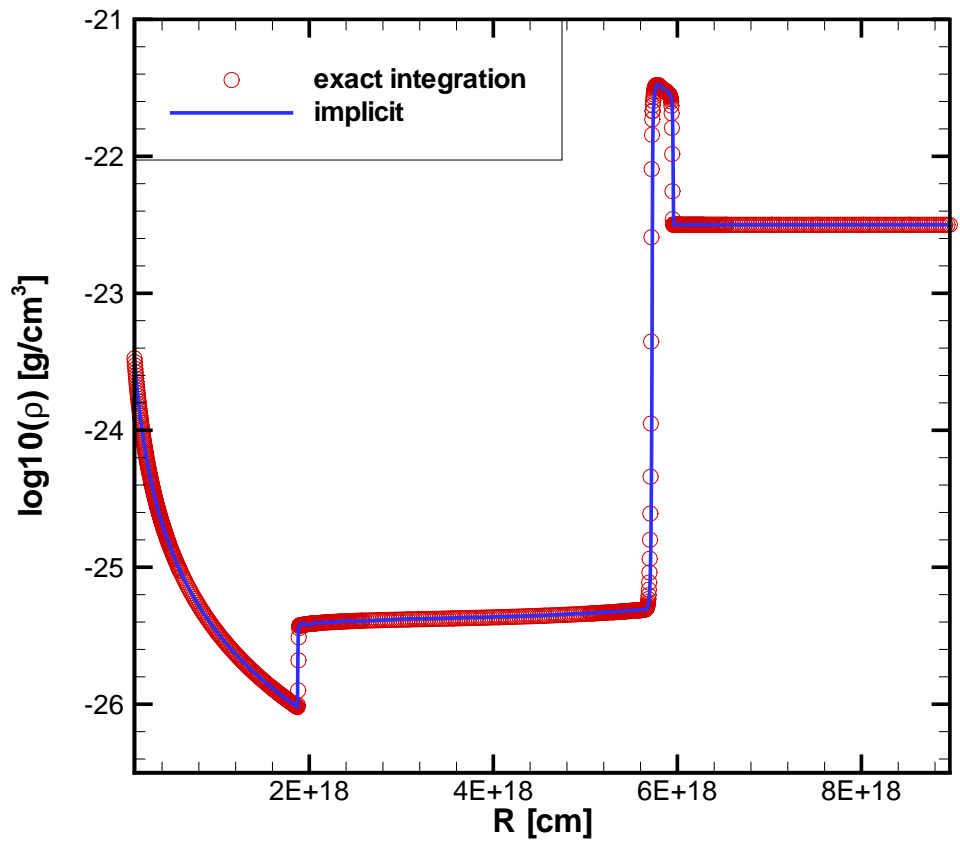}}      
\caption{Similar to fig.~\ref{fig:2coolmethods}. This time the exact integration method for the cooling is compared to a semi-implicit scheme. Again the results are identical.}
 \label{fig:2coolmethods2}
\end{figure*}

\begin{figure*}
 \centering
\resizebox{\hsize}{!}{\includegraphics[width=0.95\textwidth]{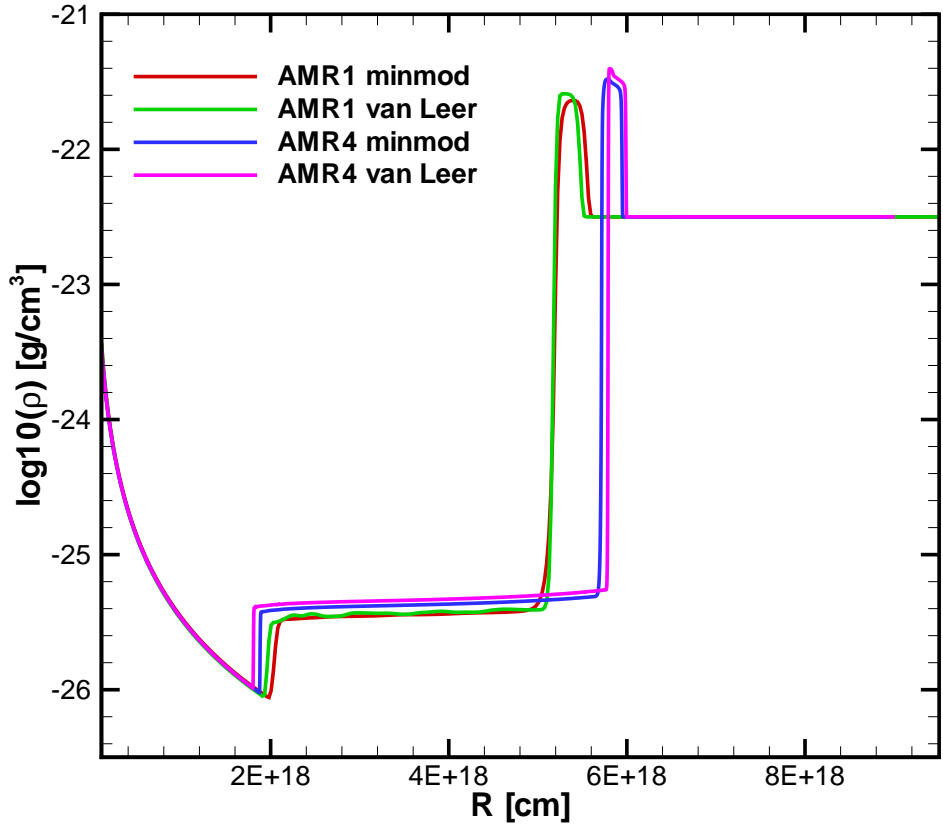}}      
\caption{The same point in time as in figs.~\ref{fig:nocool}, \ref{fig:cool1} and \ref{fig:2coolmethods} for four different simulations: Two using the van Leer flux-limiter method and two using minmod. The van Leer method is better at resolving the shell, but at low resolution shows some artificial oscillation inside the shocked wind bubble ($2\times10^{18}<R<5\times10^{18}$).}
 \label{fig:2limitmethods}
\end{figure*}

\begin{figure*}
 \centering
\resizebox{\hsize}{!}{\includegraphics[width=0.95\textwidth]{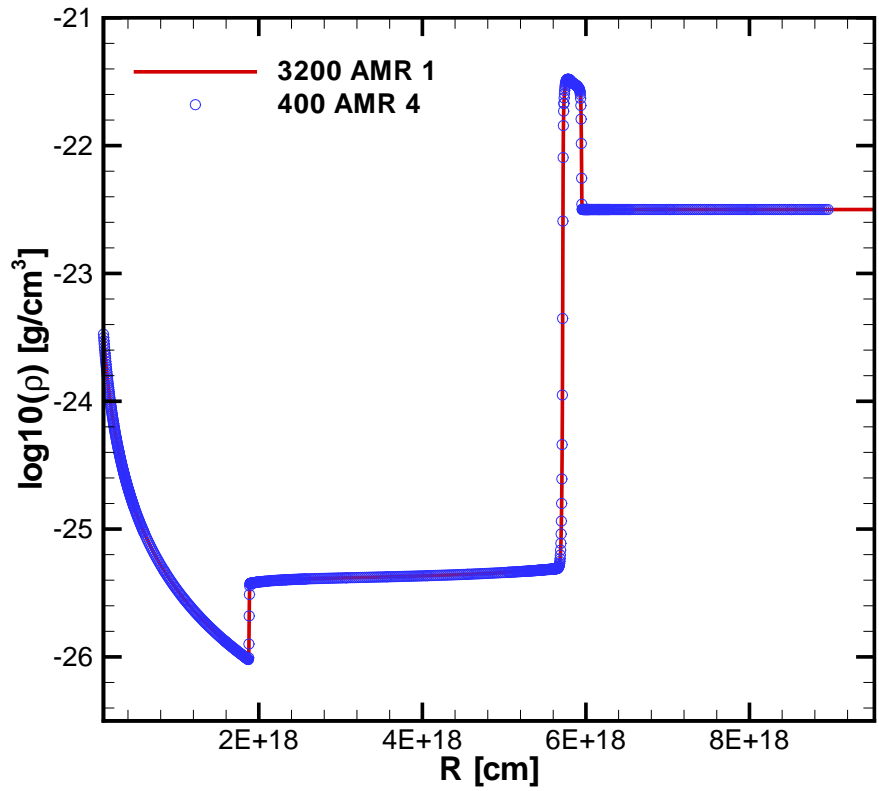}}      
\caption{The same point in time as in figs.~\ref{fig:2coolmethods} through \ref{fig:2limitmethods} for two simulations. One with a 4 level adaptive mesh grid and one with an equal resolution, but a fixed grid. The results overlap completely, showing that the adaptive mesh does not introduce numerical errors.}
\label{fig:amr}
\end{figure*}

\begin{figure*}
 \centering
\resizebox{\hsize}{!}{\includegraphics[width=0.95\textwidth]{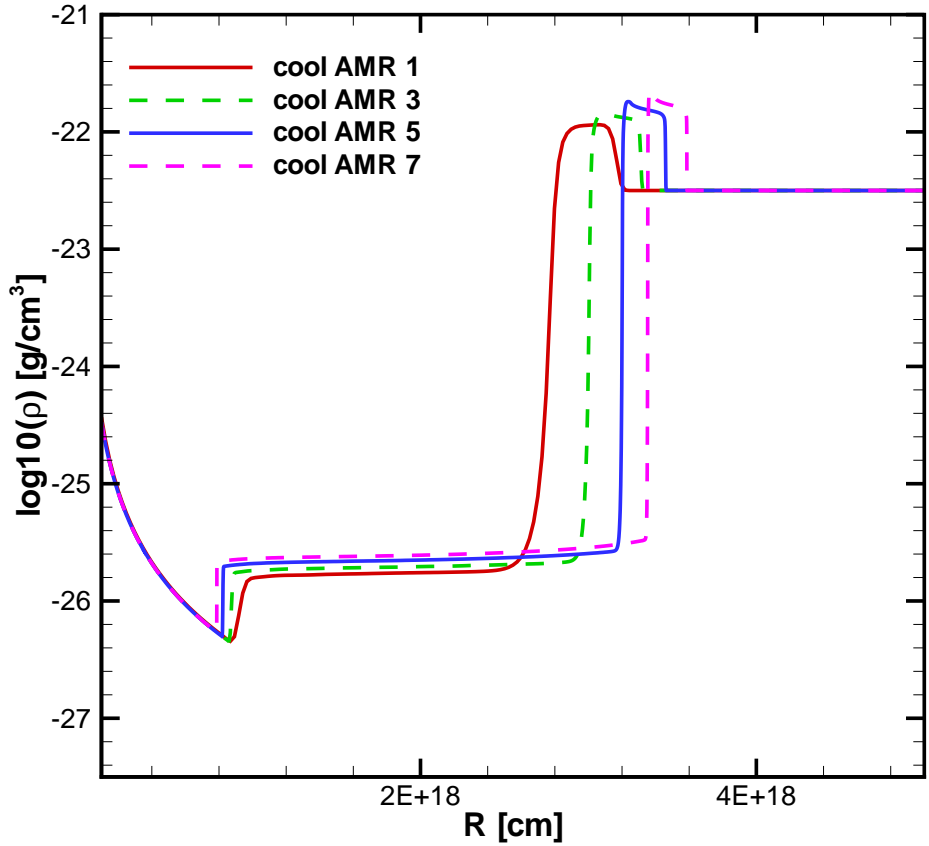}}      
\caption{The same point in time as in figs.~\ref{fig:2coolmethods} through \ref{fig:amr} for 4 simulations with a lower mass loss rate and different levels of refinement. The shell has not progressed as far (note the different scale on the horizontal axis) and is thicker. Still, the need for refinement remains as the internal structure of the shell only becomes visible at high resolution.}
\label{fig:lowmdot}
\end{figure*}

\subsection{Comparing methods}
\subsubsection{Different cooling methods}
To make sure that the exact integration method from eq.~\ref{eq:cexact} is equivalent to the treatments given by eqs.~\ref{eq:cexplicit} and \ref{eq:cimplicit}, we compare the results at the same point in time for both the AMR4 simulation with exact integration of the cooling curve and identical simulations using fully explicit and semi-implicit radiative cooling.
For the explicit method the timestep was limited so that $\delta e_{\mathrm cool} < 0.1 e_T$, where $e_T=p/(\gamma-1)$ is the internal energy density of the gas and $\delta e_{\mathrm cool}$ is the change in energy density due to radiative cooling. 
The result, shown in fig.~\ref{fig:2coolmethods}, illustrates clearly that there is no noticeable difference in the result at all. 
Similarly, fig.~\ref{fig:2coolmethods2} shows the results of the semi-implicit versus the exact integration method: again, the results are identical. 
In this particular physical problem, all three methods take approximately equally long to compute. 
The explicit method has the simplest calculation, but occasionally shorter timesteps are necessary,   
the exact integration method has a longer initial calculation before the actual simulation starts; and the implicit method requires multiple iterations per timestep. 
Using one of the Vic3 supercomputer's double quad-core Xeon 5560 nodes (2.8Ghz CPUs and 24GB of RAM) the total computation time for each method is given in Table~2. 
The main advantage of the exact integration method for this particular simulation is its reliability.  
For the explicit method the maximum allowed $\delta e_{\mathrm cool}$ must be chosen in advance and such a choice is by necessity somewhat arbitrary. 
One should actually run multiple simulations to check that the timesteps are small enough. 
The implicit method is vulnerable to potential instability, due to the particularly complex shape of the cooling curve $\Lambda(T)$ \citep{T09}.

\subsubsection{Flux-limiter influence}
As an alternative to the rather diffusive `minmod' flux-limiter we also tested our cooling routine combined with a `van Leer' \citep{L74} flux-limiter method. Comparing the results for both the AMR~1 and AMR~4 simulations, an effect becomes apparent (See Fig.~\ref{fig:2limitmethods}). 
For the lowest resolution the sharper `van Leer' limiter yields a result where the shell itself is better resolved than with the `minmod' scheme, but the total bubble is actually somewhat smaller. 
This is due to spurious oscillations in the shocked wind bubble ($2\times10^{18}<R<5\times10^{18}$). 
For the higher resolution, the van Leer method yields a result that is clearly much better than the minmod limiter. 
The van Leer AMR~4 result is then reminiscent of a minmod simulation with at least one more level of refinement.

\subsubsection{Adaptive mesh versus fixed grid}
To test the effectiveness of the adaptive mesh refinement, we compare the AMR~4 result with a fixed grid simulation with a resolution equal to the maximum resolution of the adaptive mesh, corresponding to 3200 grid points. 
The result is shown in fig.~\ref{fig:amr}, and appear indistinguishable. 
From the computation time values in Table~2 it is also clear that the adaptive mesh scheme saves a considerable amount of time. 
This is especially important for more realistic multi-D simulations.

\subsubsection{Weaker shocks}
In fig.~\ref{fig:lowmdot} we show the behaviour of a much weaker shock. For these simulations the mass loss rate of the wind has been reduced to $10^{-8}\, \mathrm{M}_\odot$/yr, reminiscent of B-type stars (between 2 and  15~$\mathrm{M}_\odot$), rather than O-type stars. Due to the lower velocity the shock temperatures are much lower ($T\sim v^2$), decreasing the effect of the radiative cooling. 
As a result, the swept-up shell which moves at a lower velocity, becomes much more extended. Still, the solution does not converge until a relatively high resolution is reached. Although the wider shell is easier to resolve, the weaker shock will in the earlier stages be more radiative (there is less energy available to radiate away, so it takes longer to form a shocked wind layer.) This in turn will lead to a longer initial phase in which the reaction is almost purely radiative, which is far more difficult to resolve. Therefore, although in the later stages the simulation will need less resolution, it starts with a larger error. Furthermore, although the temperatures in general are lower, the radiative cooling instability still occurs, which creates a necessity for a high resolution to resolve the internal structure of the shell. 

\begin{figure*}
 \centering
\resizebox{\hsize}{!}{\includegraphics[width=0.95\textwidth,angle=-90]{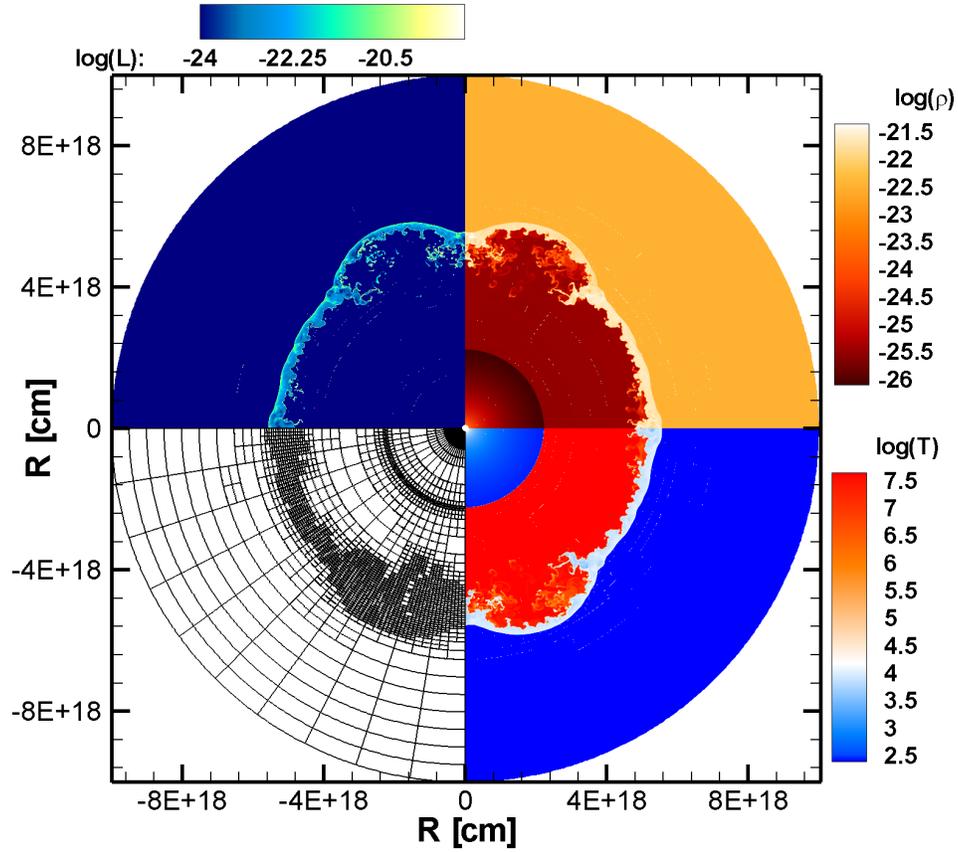}}      
\caption{The result of a 2-dimensional simulation, showing (starting upper left and moving clockwise) the luminosity, density, temperature and grid-structure. All units are in cgs format. The structure of the shell is quite complicated due to Rayleigh-Taylor instability. The adaptive mesh refinement has increased the grid resolution around the areas with steep density jumps (The shell and the reverse shock). The luminosity, which is strongly dependent on the density, only shows the shell itself.}
 \label{fig:cool2D}
\end{figure*}

\subsection{Multi-D}
\label{sec-2D}
The result of our 2-D simulation is shown in Fig.~\ref{fig:cool2D}, which shows the density, temperature and luminosity due to radiative cooling as well as the refinement of the grid. 
For this simulation we use a van Leer flux-limiter, which gives a better performance at lower resolution (see fig.~\ref{fig:2limitmethods}), and we allow a maximum of 4 levels of refinement. 
In two dimensions the shell is not a perfect spherical shape. Rather, it shows considerable structure. This is primarily due to Rayleigh-Taylor instability, induced by the fact that a low density bubble ($2\times 10^{18}<R<5\times 10^{18}$) expands into a much denser medium. 
The luminosity plot only shows the shell, owing to the strong density dependence of the radiative cooling (see eq.~\ref{eq:decool}). 
The high temperature bubble of shocked wind material barely radiates on account of its low density. 
The adaptive mesh has refined mostly around the shell and the reverse shock. 
We stress that the plot does not show the actual gridcells, these would be too small to see. What is shown is the fixed size subgrids, each consisting of 20$\times$20 gridcells.

\begin{table*}
\label{tab:time}
\caption{Computation time in CPU seconds}
\begin{tabular}{p{0.4\linewidth}lll}
method                       & total time (s)  & Startup time (s)  & I/O (s)               \\
\hline
exact integration                                              &  232.925         & 0.081                & 0.394 \\
explicit                                                              &  243.365         & 0.013                & 0.444 \\
semi-implicit                                                     &  253.205         & 0.038                & 0.699 \\
exact integration (3200 points, no AMR)          &  509.003         & 0.072                & 0.829 \\
\hline
\end{tabular}
\end{table*}

\section{Discussion of astrophysical relevance}
\label{sec-discussion}
Our results have shown the increased need for high resolution caused by the addition of radiative cooling to a numerical simulation. 
Since astrophysics usually deals with phenomena on very large scales, adaptive mesh refinement becomes a prerequisite in order to keep the total size of the simulation to a manageable level. 
Of course, this particular test-case does not cover all possible physical processes that contribute to an astrophysical problem. 
E.g. we have neglected the issue of an existing radiation field that may heat the local gas. 
Nor did we include thermal conduction, or magnetic fields. 
In reality, all these contribute, but it falls outside the scope of this paper to address them all. 
Nevertheless, we can make qualitative predictions of their influence on the final result.

\subsection{Magnetic fields}
Magnetic fields play an important role in the evolution of the circumstellar medium. 
Not only do magnetic fields exist in the interstellar gas, but the stars themselves can have powerfull magnetic fields that influence the characteristics of the stellar wind. 
The magnetic field strength in the interstellar medium is known to vary from $\sim 1\mu$G in the general interstellar medium \citep{B09, Jetal10} to 0.1-6~mG~\citep{CC07} in some massive star forming regions.  As such, the magnetic field energy density is generally low, although this can increase if the gas is swept up by the stellar wind. 
Much will depend on the alignment of the magnetic field lines with the motion of the swept up shell. 
If the field lines run parallel with the direction in which the shell moves, the influence of the field on the general morphology of the gas is likely negligible. 
However, should the field lines run parallel with the shell, the local magnetic field energy density increases as the gas is compressed. This causes the shell to expand as the magnetic field counteracts the pressure, resulting in a lower density of the shell. This acts to decrease the effect of radiative cooling. 

The magnetic field of the star itself is unlikely to be relevant, although some massive stars have strong magnetic fields.
Simulations by \citet{UOT08, UOT09} show that far from the star ($\gtrsim 10$ stellar radii) the magnetic field lines 
are torn open by the wind. This results in a monopolar field structure, where the field lines become parallel with the direction of the wind. 
Therefore, the stellar field will not influence the kinetics of the circumstellar gas at large distances.

\subsection{Radiation fields}
The absorption of high energy photons from an outside source (such as nearby stars) can influence the morphology of the circumstellar medium and photo-ionize the circumstellar gas. 
The effect of photo-ionization was described in a simple 1D model by \citet{VLG05, VLG07} 
and 2D simulations of similar problems were run by \citet{FHY03,FHY06}.
The main effect is the increased temperature of the material into which the stellar winds expands. 
This raises the sound speed and depending on the strength of the wind, the expansion speed of the swept-up shell 
can drop below this sound speed. Then, the forward shock is lost and the transition between expanding shocked wind and surrounding (ionized) medium becomes a contact discontinuity rather than a shock \citep{VLG05}. At the outer edge of the photo-ionized region, the thermal pressure of the photo-ionized gas will still cause a shock to form 
as the ionized gas expands into the cold surrounding medium. 
On the other hand, if the wind is strong enough to maintain a supersonic expansion, the result will be almost identical to the situation without photo-ionization \citep{VLG07}. 
Photo-ionization can also increase the instability of a partially ionized shell, as local photo-ionized regions have a much higher temperature than their surroundings and will expand \citep{GLRF99}. However, the shells we find already have a temperature of about 10,000~K (the ionization temperature of hydrogen) due to collisional heating. Therefore, it is unlikely that ionization due to the presence of radiation will have a large effect on the structure of the shell. 

\subsection{Going to 3-D}
As shown in section \ref{sec-2D}, the circumstellar shell is highly susceptible to instabilities in more than 1 spatial dimension.  
The full implications of thin-shell instabilities in 3D have not yet been explored in great detail, and form part of our ongoing efforts. 
Generally though, instabilities tend to form quicker in 3D than in 2D \citep[e.g.][]{YTDR01}, quickly producing higher density contrasts and creating more complicated structures. 
This can only increase the effects described in this paper, as higher densities lead inevitably to greater radiative cooling, which in turn will 
increase the density as the cooling gas is compressed.  
Once again, there is a strong need for a high resolution grid and therefore the application of AMR will be inevitable.

\section{Conclusion}
\label{sec-conclusion}
In order to properly resolve the thin, high density shells resulting from radiative cooling, astrophysical gas dynamics simulations need grids with high resolution. 
Since the number of gridpoints would quickly become prohibitive if this simulation was carried out on a fixed grid, 
adaptive mesh refinement can be considered a necessity for simulations of this kind. 
In 2-D (and 3-D) the adaptive mesh becomes even more crucial, since the number of gridpoints is much larger to begin with and the complicated structures need high resolution to be properly resolved.
Numerically, a better result could be achieved by replacing the diffusive minmod scheme in a linear extrapolation with a higher order interpolation method, such as the piecewise parabolic method \citep{Fetal00, MPB05}. This would allow the code to reach a more accurate solution at lower resolution. 

Finally, this particular simulation has rather high densities and shock temperature due to the high mass loss rate of the stellar wind. 
This in turns increases the effect of the radiative cooling, due to its strong density dependence.  
Weaker shocks can be equally problematic to resolve due to their more radiative nature in the inital collision.
This could perhaps be mitigated by starting the simulation gradually, rather than have the full strength wind interact directly with the stationary gas. Unfortunately, any simulation then depends on parameters used to control the transition. 
Also, high density regions will still be compressed due to radiative cooling, which means that high resolution grids will be needed.

We also stress the actual scale that these simulations cover. 409\,600 gridcells along one dimension may seem excessive, 
but since the whole grid covers a range of $10^{19}$ cm, this means that each gridcell has a cross-section of about 35 solar radii.
Of course, one can wonder whether a proper resolution of the shell is really necessary.
For one thing, typical astrophysical problems involve significant uncertainties in the input parameters, which dwarf the possible effect of the resolution. 
However, for anyone working on such simulations it is prerequisite to be aware of this particular problem and consider how it may affect scientific results. A viable alternative to AMR grids as used here, is the use of moving grids, which have recently been used in pure 2D Euler flows in~\citet{vandam}. The effect of radiative losses as emphasized here still need to be evaluated for moving grid simulations. 

\section{Acknowledgments}
A.J.v.M.\ acknowledges support from NSF grant AST-0507581, from the
FWO, grant G.0277.08 and K.U.Leuven GOA/09/009. 
We thank dr.~G.~Mellema for the use of the cooling table from \citet{ML02}. 
A.J.v.M. thanks dr.~Z.~Meliani for his help with setting up the MPI-AMRVAC code. 
All simulations were done on the Vic3 HPC cluster at K.U. Leuven. 


\end{document}